\documentclass[aps,showpacs,preprintnumbers,amsmath,amssymb]{revtex4}
\usepackage{graphicx}
\usepackage{epstopdf}
\usepackage{latexsym}
\usepackage{amssymb}
\usepackage{amsmath}
\usepackage{color}
\usepackage{float}
\usepackage{mathrsfs}
\usepackage[center]{subfigure}
\usepackage{dcolumn}
\usepackage[dvips]{epsfig}

\usepackage{graphicx}
\usepackage{epstopdf}
\usepackage{latexsym}
\usepackage{amssymb}
\usepackage{amsmath}
\usepackage{color}
\usepackage{mathrsfs}
\usepackage[center]{subfigure}
\usepackage{ifpdf}
\usepackage{epstopdf}

\begin{document}
\baselineskip=0.8 cm
\title{{\bf Gravitational lensing by a black hole with torsion }}

\author{Lu Zhang$^{1,2}$, Songbai Chen$^{1,2,3}$\footnote{Corresponding author: csb3752@hunnu.edu.cn}, Jiliang
Jing$^{1,2,3}$ \footnote{jljing@hunnu.edu.cn}}
%\email{csb3752@163.com}

\affiliation{$^{\textit{1}}$Institute of Physics and Department of Physics, Hunan
Normal University,  Changsha, Hunan 410081, People's Republic of
China \\ $^{\textit{2}}$Key Laboratory of Low Dimensional Quantum Structures \\
and Quantum Control of Ministry of Education, Hunan Normal
University, Changsha, Hunan 410081, People's Republic of China\\
$^{\textit{3}}$Synergetic Innovation Center for Quantum Effects and Applications,
Hunan Normal University, Changsha, Hunan 410081, People's Republic
of China}

\vspace*{0.2cm}
\begin{abstract}
\baselineskip=0.6 cm
\begin{center}
{\bf Abstract}
\end{center}

In this paper we have investigated the gravitational lensing in a spherically symmetric spacetime with torsion in the generalized Einstein-Cartan-Kibble-Sciama (ECKS) theory of gravity by considering higher order terms. The torsion parameters change the spacetime structure which affects the photon sphere, the deflection angle and the strong gravitational
lensing. The condition of existence of horizons is not inconsistent with that of the photon sphere. Especially,
there exists a novel case in which there is horizon but no photon sphere for the considered spacetime. In this special case, the deflection angle of the light ray near the event horizon also diverges logarithmically, but the coefficients in the strong-field limit are different from those in the cases with photon sphere. Moreover, in the far-field limit, we find that the deflection angle for certain torsion parameters approaches zero from the negative side, which is different from those in the usual spacetimes.
\end{abstract}

 \maketitle
\newpage
\vspace*{0.2cm}
\section{Introduction}

 General relativity is considered probably the most beautiful theory of gravity at present
 in which the gravitational interaction is described in a remarkable way as
 a purely geometrical effect of the spacetime. It has successfully passed a series of observational and experimental tests \cite{t3} and becomes a fundamental theoretical setting for the modern astrophysics and cosmology.
However, there exist a lot of efforts focusing on the study of other alternative theories of gravity since in Einstein theory of gravity an important ingredient is missing which may account for the observed accelerating expansion of the current Universe \cite{t5,t501,t51,t6,t61}.
Torsion is one of such ingredients which vanish in Einstein theory. In the theories with torsion, the presence of torsion will change the character of the gravitational interaction because that the gravitational field is described not only by spacetime metric,
but also by the torsion field.

The ECKS theory of gravity \cite{Cartan,Kibble} is a natural extension of Einstein's general relativity
in which  the curvature and the torsion are assumed to
couple with the energy and momentum and the intrinsic angular momentum of matter, respectively. Due to this coupling between matter and the affine connection, the
intrinsic angular momentum of matter can be acted as a source of torsion.
At the region with extremely large densities including the interior of black holes and the very early stage of Universe, the minimal spinor-torsion coupling leads to a gravitational repulsion effect and avoids the formation
of spacetime singularities \cite{avoid,avoid1,avoid2,avoid3}. At the low densities region, the contribution from torsion to the Einstein equations is negligibly small so that the ECSK theory and Einstein's general relativity give indistinguishable predictions. Moreover, the torsion field outside of matter distribution vanishes in the ECSK theory since the torsion equation is an algebraic constraint rather than a partial differential equation and then the torsion field does not propagate as a wave in the spacetime.

It is possible to construct a generalized ECSK theory describing a dynamical torsion
field by adding higher order corrections in Lagrangian \cite{gK1,gK2}. In this generalized ECSK theory,
the spacetime torsion and curvature couple each other so that both equations of motion for
spacetime torsion and curvature are dynamical equations which ensures that the spacetime torsion can propagate in the spacetime even in the absence of spin of matter. Recently, the effects of higher order terms from spacetime curvature and torsion has been studied and some non-trivial static vacuum solutions \cite{sbh1} are obtained in this generalized ECSK theory. This non-trivial solutions describes the spacetimes with special structures, which is useful for
detecting the effects originating from spacetime torsion within the gravitational interactions.

Gravitational lensing is such a phenomenon caused
by the deflection of light rays in a gravitational field.
Like a natural telescope, gravitational lensing can provide us
not only some important signatures about black holes in the
Universe but also profound verification of alternative theories
of gravity in their strong field regime \cite{Ein1,Darwin,KS1,KS2,KS4, VB1,VB2,VB202,Gyulchev,Gyulchev1,Fritt,Bozza1,Eirc1,whisk,
Bhad1,Song1,Song2,TSa1,AnAv,agl1,agl101,gr1,gr2,gr201,gr3,gr4,gr401}. Thus,
the gravitational lensing is regarded as a powerful
indicator of the physical nature of the central celestial
objects and then has been studied extensively in various theories of gravity.
In this paper, we will focus on the gravitational
lensing by the spacetime with torsion obtained in this generalized ECSK theory and probe the effects originating from spacetime torsion.

The paper is organized as follows. In Sec. II, we review briefly the metric of spacetime with torsion and then analyze the deflection angles for light ray propagating in this background. In Sec. III, we investigate strong gravitational lensing near the photon sphere and near the horizon in the spacetime with torsion. Finally, we present a summary.

\section{The spacetime of a black hole with torsion and the deflection angle for light ray}

Firstly, let us review briefly a black hole solution with torsion in the generalized Einstein-Cartan-Kibble-Sciama (ECKS) theory of gravity, which is obtained in Ref. \cite{sbh1}. The corresponding action can be expressed as
\begin{eqnarray}
S&=&\int d^4x \sqrt{-g}\bigg[-\frac{1}{16\pi G}\bigg(\mathcal{L}+a_1\mathcal{L}_1\bigg)\bigg],
\label{action}
\end{eqnarray}
where $a_1$ is a coupling constant, $\mathcal{L}$ is Einstein-Cartan-Kibble-Sciama (ECKS) Lagrangian \cite{gK1,gK2,sbh1}
\begin{eqnarray}
\mathcal{L}&=& R+\frac{1}{4}Q_{\alpha\beta\gamma}Q^{\alpha\beta\gamma}
+\frac{1}{2}Q_{\alpha\beta\gamma}Q^{\beta\alpha\gamma}+
Q^{\alpha\;\beta}_{\;\alpha}Q^{\gamma}_{\;\beta\gamma}+2Q^{\alpha\;\beta}_{\;\alpha;\beta},
\end{eqnarray}
and $\mathcal{L}_1$ is
\begin{eqnarray}
\mathcal{L}_1&=& RQ_{\alpha\beta\gamma}Q^{\alpha\beta\gamma}
+\frac{1}{4}Q_{\alpha\beta\gamma}Q^{\alpha\beta\gamma}\bigg[Q^{\delta\epsilon\eta}
(2Q_{\epsilon\delta\eta}+Q_{\delta\epsilon\eta})+8Q^{\delta\;\epsilon}_{\;\delta;\epsilon}\bigg]+
Q^{\alpha\;\beta}_{\;\alpha}Q^{\gamma}_{\;\beta\gamma}Q^{\delta\epsilon\eta}Q_{\delta\epsilon\eta}.
\end{eqnarray}
Here $R$ is Ricci Scalar and $Q^{\alpha}_{\;\mu\nu}$ is the spacetime torsion, which is the antisymmetric part of the general affine connection $\tilde{\Gamma}^{\alpha}_{\;\mu\nu}$,
\begin{eqnarray}
Q^{\alpha}_{\;\mu\nu}=\tilde{\Gamma}^{\alpha}_{\;\mu\nu}-\tilde{\Gamma}^{\alpha}_{\;\nu\mu}.
\end{eqnarray}
The affine connection $\tilde{\Gamma}^{\alpha}_{\;\mu\nu}$ is related to the Levi-Civita Christoffel connection $\Gamma^{\alpha}_{\;\mu\nu}$ by
\begin{eqnarray}
\tilde{\Gamma}^{\alpha}_{\;\mu\nu}=\Gamma^{\alpha}_{\;\mu\nu}+K^{\alpha}_{\;\mu\nu},
\end{eqnarray}
where $K^{\alpha}_{\;\mu\nu}$ is the contorsion tensor with a form
\begin{eqnarray}
K^{\alpha}_{\;\mu\nu}=\frac{1}{2}\bigg[Q^{\alpha}_{\;\mu\nu}-Q^{\;\alpha}_{\mu\;\nu}
-Q^{\;\alpha}_{\nu\;\mu}\bigg].
\end{eqnarray}
This action admits  a static spherically symmetric vacuum solution with the line element as followed \cite{sbh1}
\begin{eqnarray}
ds^2 = -H(r) dt^2 + \frac{dr^2}{F(r)} + r^2 (d\theta^2
+\sin^2\theta d\phi^2), \label{metr}
\end{eqnarray}
with
\begin{eqnarray}
\label{sol1}
F(r)&=&1-\frac{c_1}{c_2\sqrt{r}}+\frac{6c_3c^2_2-9c^2_1\ln(3c_1-2c_2\sqrt{r})}{6c^2_2r},
\nonumber\\
H(r)&=&\bigg(1-\frac{2c_1}{c_2\sqrt{r}}\bigg)^2,
\end{eqnarray}
where $c_1$, $c_2$ and $c_3$ are the integration constants related to torsion.
Due to the existence of logarithmic term in function $F(r)$, we must require that the factor $3c_1-2c_2\sqrt{r}$ is positive at the physical region. As $c_2>0$, one can find that the factor $3c_1-2c_2\sqrt{r}$ is always negative in the case $c_1<0$, while for the case $c_1>0$, there exists a constraint for the polar coordinate $r$, i.e., $\sqrt{r}<\frac{3c_1}{2c_2}$ and the function $F(r)$ is divergent  as $r$ approaches this limit. Therefore, we here focus only on the case $c_2<0$ rather than the case $c_2>0$. Moreover, in the solution (\ref{metr}), the coupling constant $a_1$ is set to $a_1=\frac{2}{9c^2_2}$ so that it is asymptotical flat at spatial infinite as $c_2<0$.

For convenience, we set $c_2=-1$ in our throughout analysis. The position of event horizon of the black hole is defined by $F(r)=0$, which is obtained numerically since the emergence of logarithmic term. We find that there exist a critical value for
the existence of horizon in the spacetime (\ref{metr})
\begin{eqnarray}
 c_{crit}=\left\{\begin{array}{cc}
 \frac{3c^2_1}{2}\ln(3c_1),& \;\;\;\;\; c_1>0, \nonumber\\
 \frac{c^2_1}{2}[3\ln(-c_1)-4], & \;\;\;\;\;c_1<0.
 \end{array}\right.
 \label{crti1}
\end{eqnarray}
The dependence of the existence of horizon on the integration constants $c_1$ and $c_3$ is plotted in Fig.(1).
\begin{figure}[ht]
\begin{center}
\includegraphics[width=7cm]{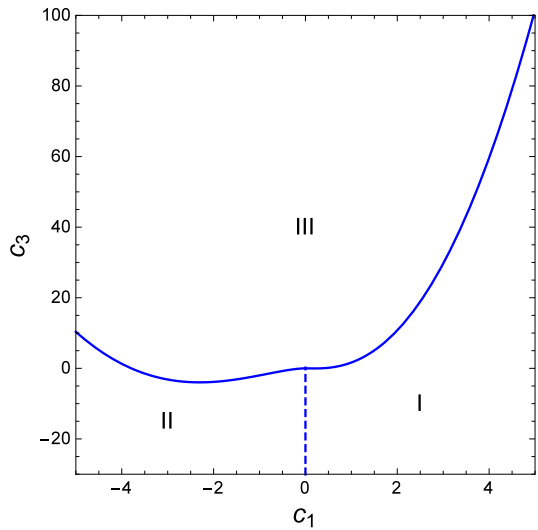}
\caption{The dependence of the existence of horizon on the integration constants $c_1$ and $c_3$. The regions $I$, $II$ and $III$ separated by curves $c_3=c_{crit}$ and $c_1=0$ in the panel are corresponded to the cases with single horizon, two horizons and no any horizon,
respectively.  Here we set $c_2=-1$.}
\end{center}
\end{figure}
As $c_3>c_{crit}$, there is no horizon and the metric describes the geometry of a naked singularity. For the case with $c_3<c_{crit}$, we find that the black hole possesses a single horizon as $c_1>0$ and two horizons as $c_1<0$. Moreover, for the case $c_1>0$, we find that there is no solution of the equation $H(r)=0$, which means that there is no infinite redshift surface in this case. For the case with $c_1<0$, the infinite redshift surface is located at
$r_{\infty}=4c^2_1$, but inside the event horizon. Especially, in the case with horizon, the  surface gravity constant of event horizon is
\begin{eqnarray}
\kappa=\sqrt{\frac{g^{rr}}{g_{tt}}}\frac{dg_{tt}}{dr}\bigg|_{r=r_H}=
\sqrt{\frac{F(r)}{H(r)}}\frac{d H(r)}{dr}\bigg|_{r=r_H}=0.
\end{eqnarray}
Thus, this black hole solution possesses some particular spacetime properties differed from those in the Einstein's general relativity, which could make a great deal influence on the propagation of photon in the spacetime.

Let us to study the gravitational lensing in the background of a black hole spacetime with torsion (\ref{metr}). For simplicity, we here focus on the case in which both the source and the observer lie in the equatorial plane in the spacetime so that the orbit of the photon is limited on the same plane. Under this condition $\theta=\pi/2$, the reduced metric (\ref{metr}) can be expressed as
\begin{eqnarray}
ds^2=-A(r)dt^2+B(r)dr^2+C(r)d\phi^2,\label{grm}
\end{eqnarray}
with
\begin{eqnarray}
A(r)=H(r), \;\;\;\;B(r)&=&1/F(r),\;\;\;\; C(r)=r^2.
\end{eqnarray}
The null geodesics for the metric (\ref{grm}) obey
\begin{eqnarray}
\frac{dt}{d\lambda}&=&\frac{1}{A(r)},\label{u3}\\
\frac{d\phi}{d\lambda}&=&\frac{J}{C(r)},\label{u4}\\
\bigg(\frac{dr}{d\lambda}\bigg)^2&=&\frac{1}{B(r)}\bigg[\frac{1}{A(r)}
-\frac{J^2}{C(r)}\bigg].\label{cedi}
\end{eqnarray}
where $J$ is the angular momentum of the photon and $\lambda$ is an affine parameter along the null geodesics. Here, the energy of photon is set to $E=1$. In the background of a black hole spacetime with torsion (\ref{metr}), the relation
between the impact parameter $u(r_0)$ and the distance of the closest approach of the light ray $r_0$ can obtained by the conservation of the
angular momentum along the null geodesics
\begin{eqnarray}
u(r_0)=J(r_0)=\frac{r_0}{\sqrt{H(r_0)}}=\frac{r_0^{3/2}}{\sqrt{r_0}+2c_1}.
\end{eqnarray}
The photon sphere is very important for the propagation of photon in the black hole spacetime. In the background of a black hole spacetime with torsion (\ref{metr}), the radius of the photon sphere $r_{ps}$ is the largest
real root of the equation
\begin{eqnarray}
A(r)C'(r)-A'(r)C(r)=2(\sqrt{r}+2c_1)(\sqrt{r}+3c_1)=0.\label{root}
\end{eqnarray}
Obviously, there is no photon sphere as $c_1>0$ since all of the roots of above equation are negative. For the case of $c_1<0$, one can obtain that the largest
real root of equation (\ref{root}) is $r_{ps}=9c^2_1$. However, the further analysis indicates the value of $r_{ps}$ is smaller than that of the radius of the event horizon as $c_3<\frac{3}2c^2_1[\ln(-3c_1)-4]$, which means the surface $r=r_{ps}$ is inside the event horizon and there is no real photon sphere in this case.
\begin{figure}
\begin{center}
\includegraphics[width=7cm]{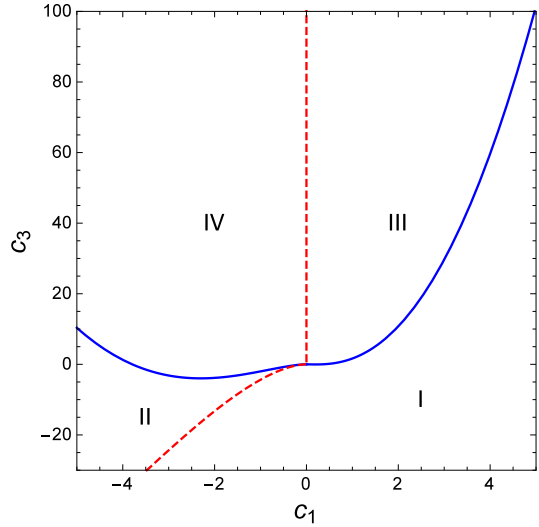}
\caption{ The boundary of the existence of horizon (blue line) and of the photon sphere (red dashed line) in the background of a black hole spacetime with torsion (\ref{metr}).  Here, we set $c_2=-1$.}
\end{center}
\end{figure}
Comparing with the previous discussion about the horizon, we find that the condition of existence of horizons is not inconsistent with the photon sphere, which is similar to that in the Konoplya-Zhidenko rotating non-Kerr spacetime \cite{Song2}. However, the situation is more complicated in the black hole spacetime with torsion (\ref{metr}). In Fig.(2),  the blue solid line and the red dashed line, respectively, correspond to the boundary of the existence of horizon and of the photon sphere, which split the whole region into four regions.
There exist both horizon and  photon sphere when the parameters ($c_1, c_3$) lie in the region $II$. This situation is similar to that of in the static black hole spacetime in the Einstein's general relativity where the photon sphere lies outside the event horizon.
Moreover, as ($c_1, c_3$) lies in the region $III$, there is no horizon and no photon sphere, which means that the singularity is completely naked in this case. However, when ($c_1, c_3$) is located in region $IV$, there is no horizon but the singularity is covered by the photon sphere. These two situations are correspond to the cases of strong naked singularity (SNS) and weakly naked singularity (WNS) \cite{KS4,Gyulchev1}, respectively. Especially, as ($c_1, c_3$) lies in the region $I$, there is horizon but no photon sphere, which is not appeared else in the existing literatures.

We are now in position to discuss the
behavior of the deflection angle for the lens described by a metric with torsion (\ref{metr}). The deflection angle for the photon coming from infinite can be expressed as
\begin{eqnarray}
\alpha(r_0)=I(r_0)-\pi,
\end{eqnarray}
where $r_0$ is the closest approach distance and $I(r_0)$ is
\cite{Ein1}
\begin{eqnarray}
I(r_0)=2\int^{\infty}_{r_0}\frac{\sqrt{B(r)}dr}{\sqrt{C(r)}
\sqrt{\frac{C(r)A(r_0)}{C(r_0)A(r)}-1}}.\label{int1}
\end{eqnarray}
\begin{figure}[ht]\label{pas22}
\begin{center}
\includegraphics[width=6cm]{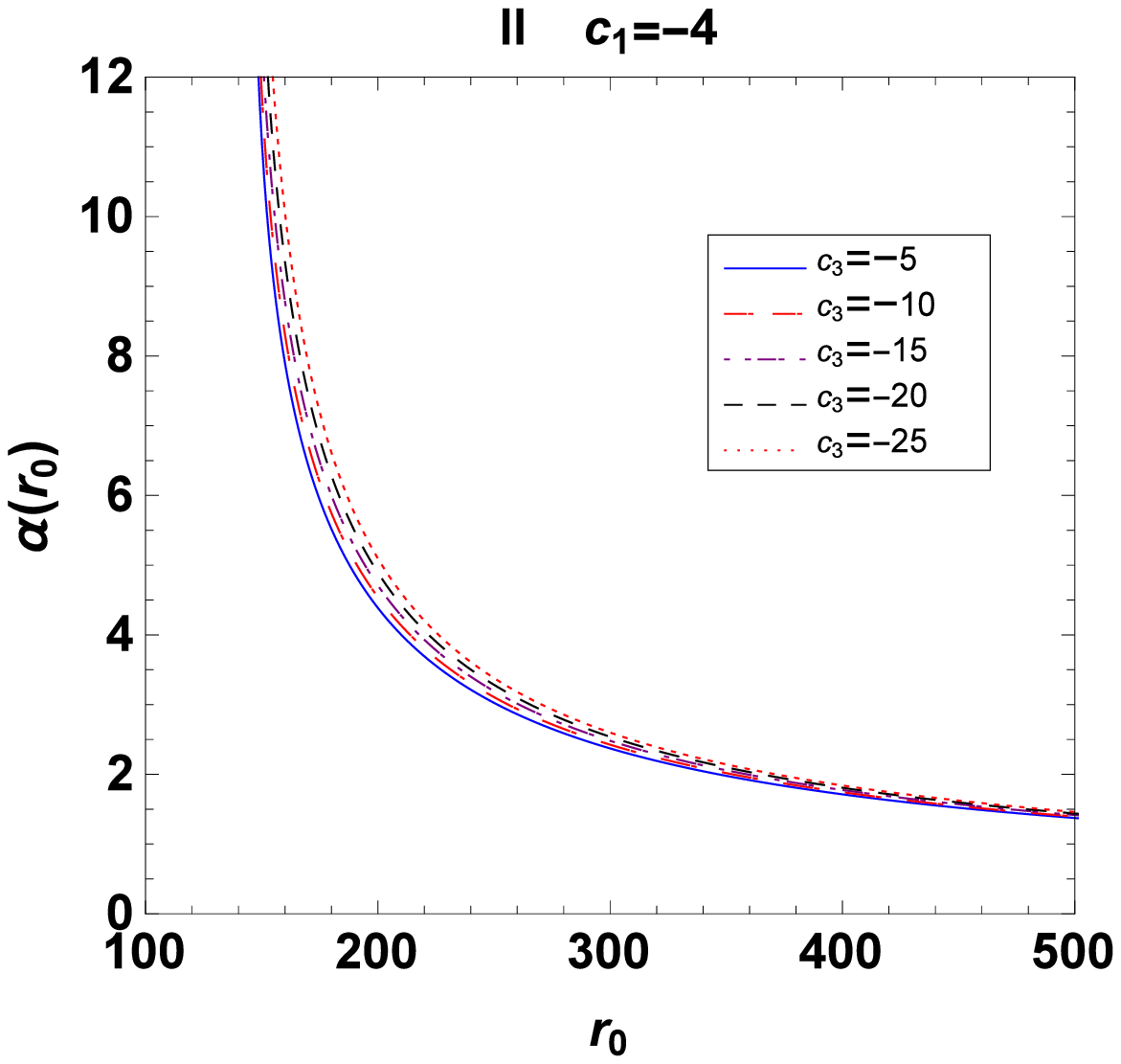}\includegraphics[width=6cm]{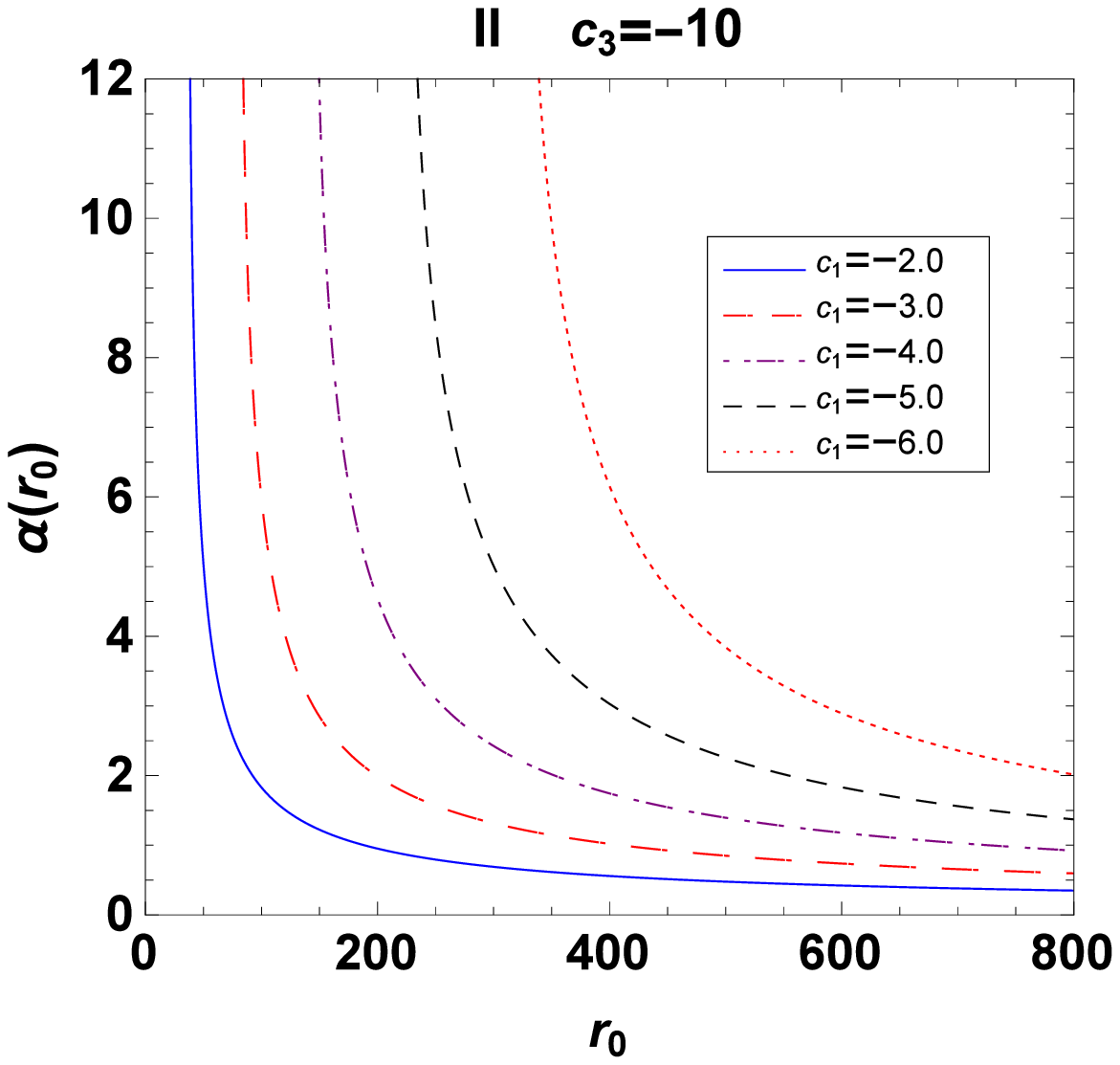}\\
\includegraphics[width=6cm]{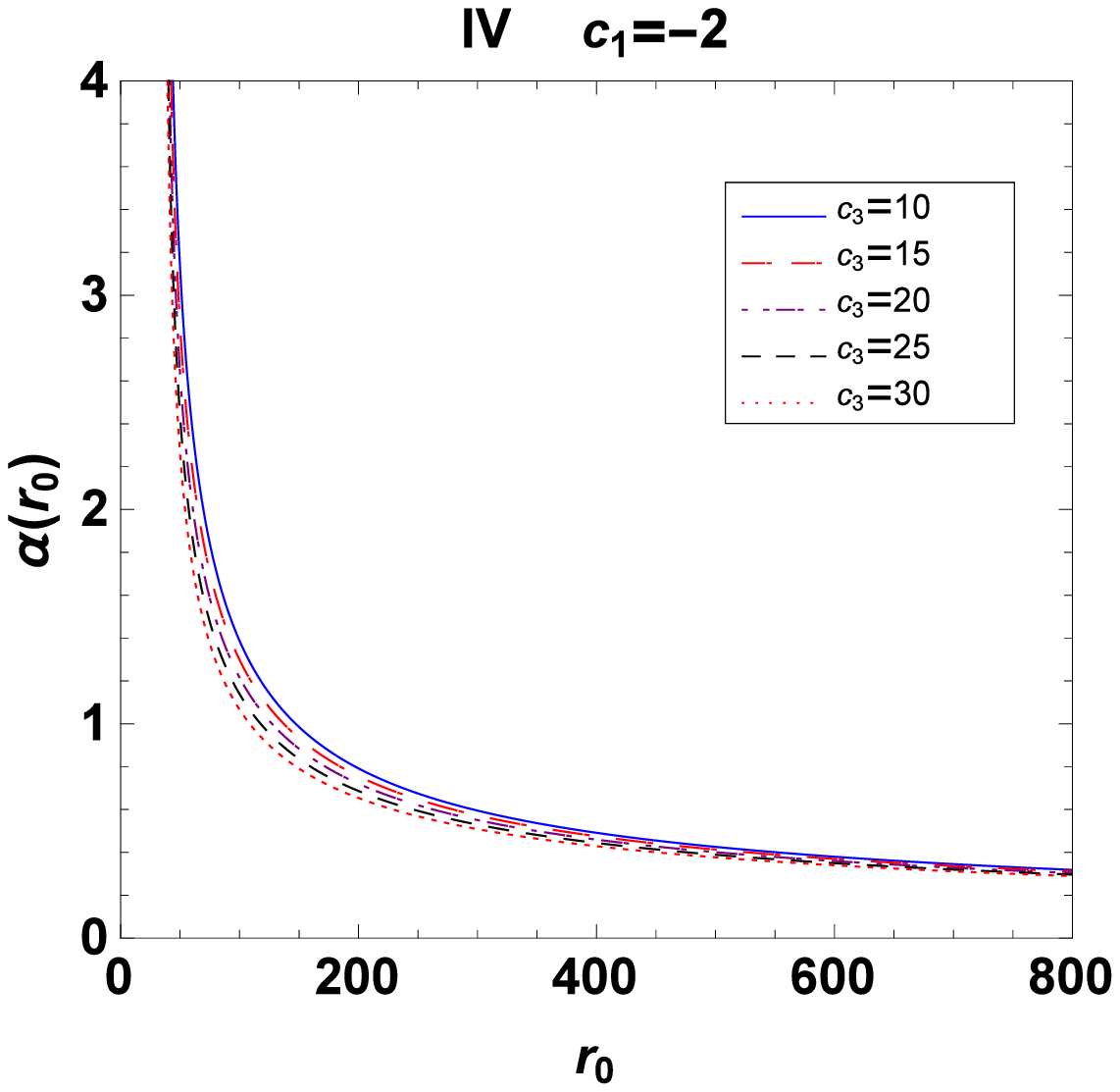}\includegraphics[width=6cm]{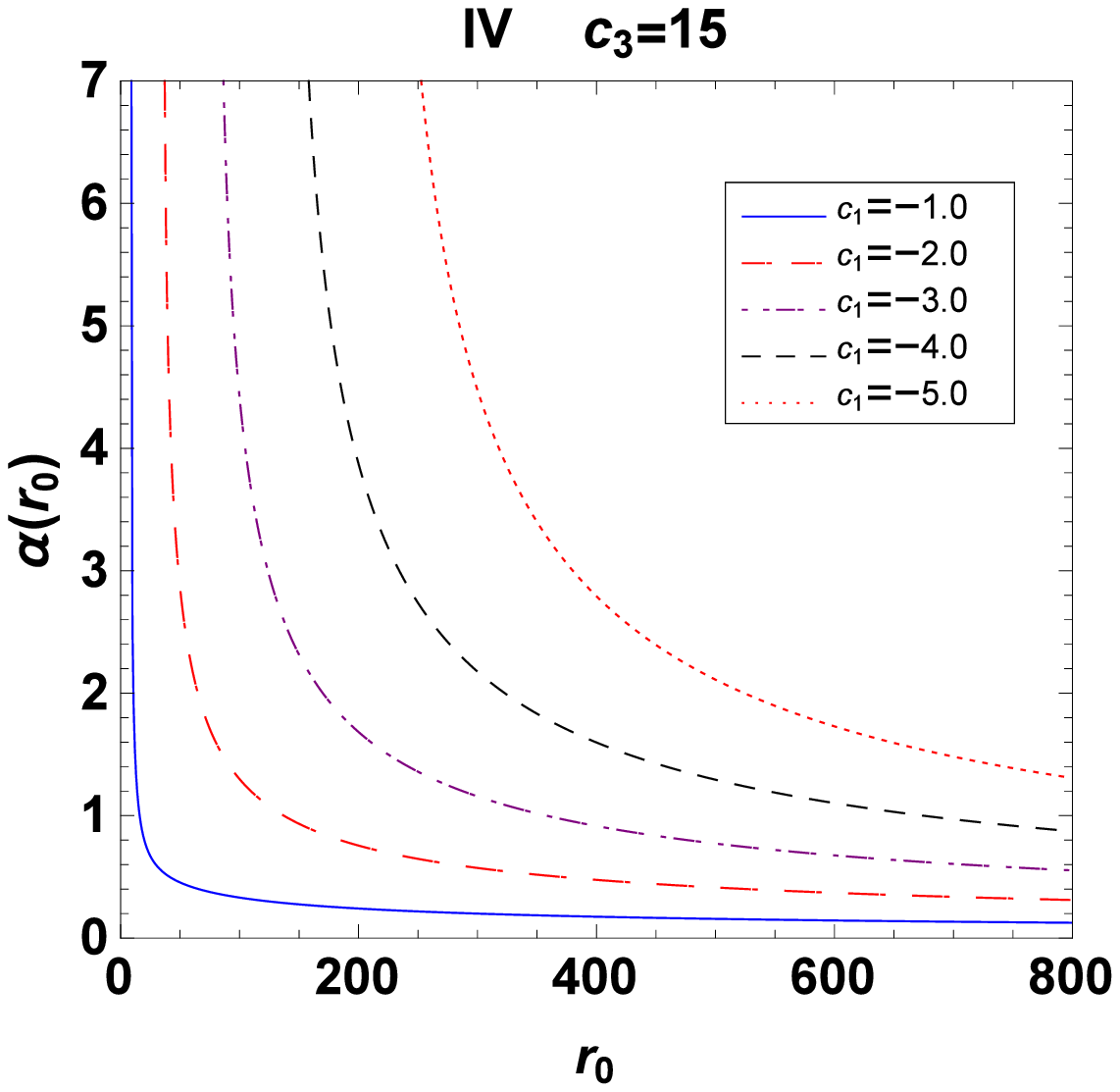}
\caption{Deflection angle $\alpha(r_0)$ as a function of the closest
distance of approach $r_0$ for the cases with photon sphere. The panels in the upper and bottom rows correspond to the cases in which the parameters ($c_1, c_3$) are located in the regions $II$ and $IV$ in Fig.(2), respectively. Here, we set $c_2=-1$.}
\end{center}
\end{figure}
\begin{figure}[ht]\label{pas3}
\begin{center}
\includegraphics[width=6cm]{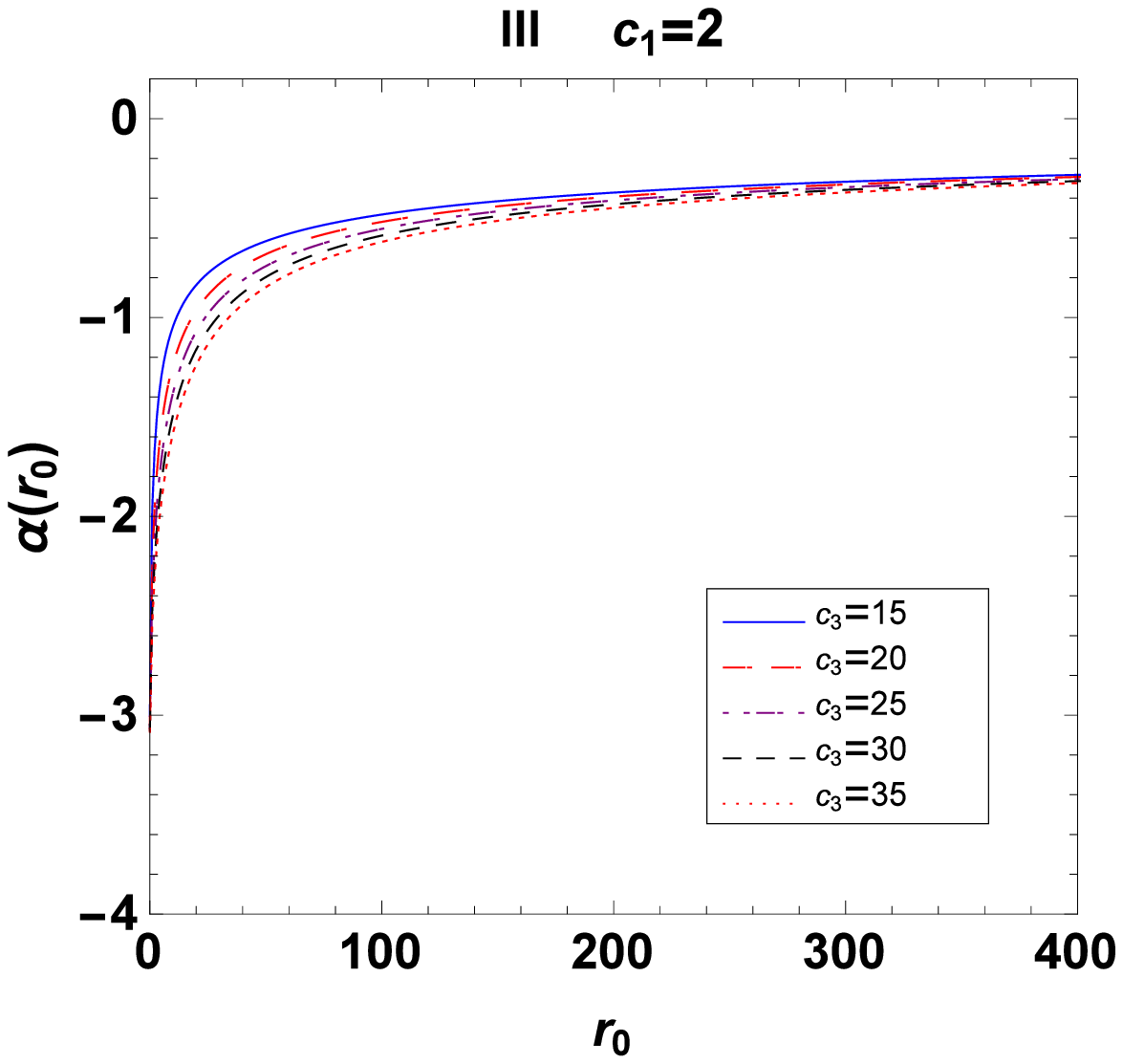}\includegraphics[width=6cm]{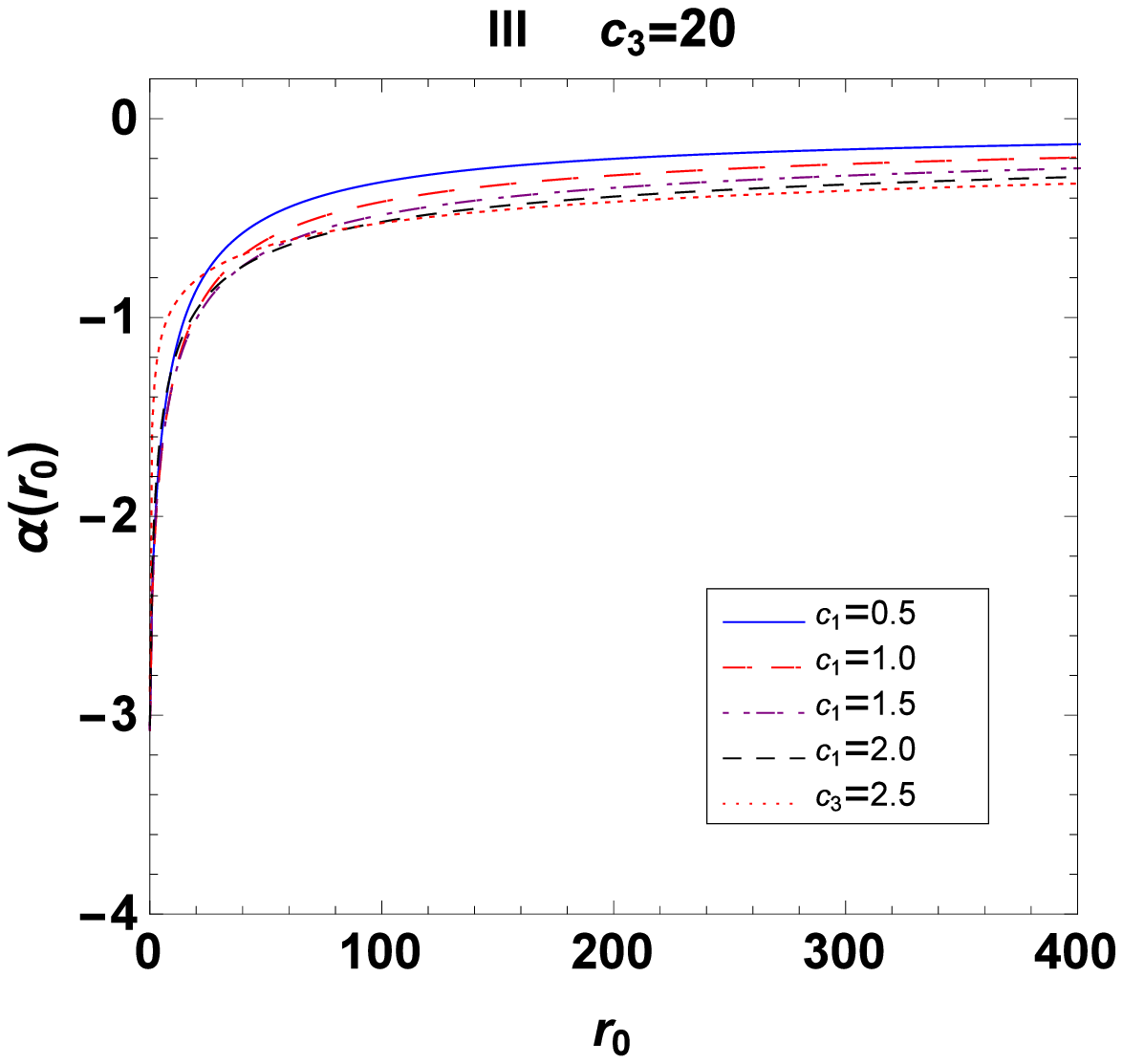}
\caption{Deflection angle $\alpha(r_0)$ as a function of the closest
distance of approach $r_0$ for the case without photon sphere and horizon in which the parameters ($c_1, c_3$) lie in the region $III$ in Fig.(2). Here, we set $c_2=-1$.}
\end{center}
\end{figure}
\begin{figure}[ht]\label{pas4}
\begin{center}
\includegraphics[width=5.2cm]{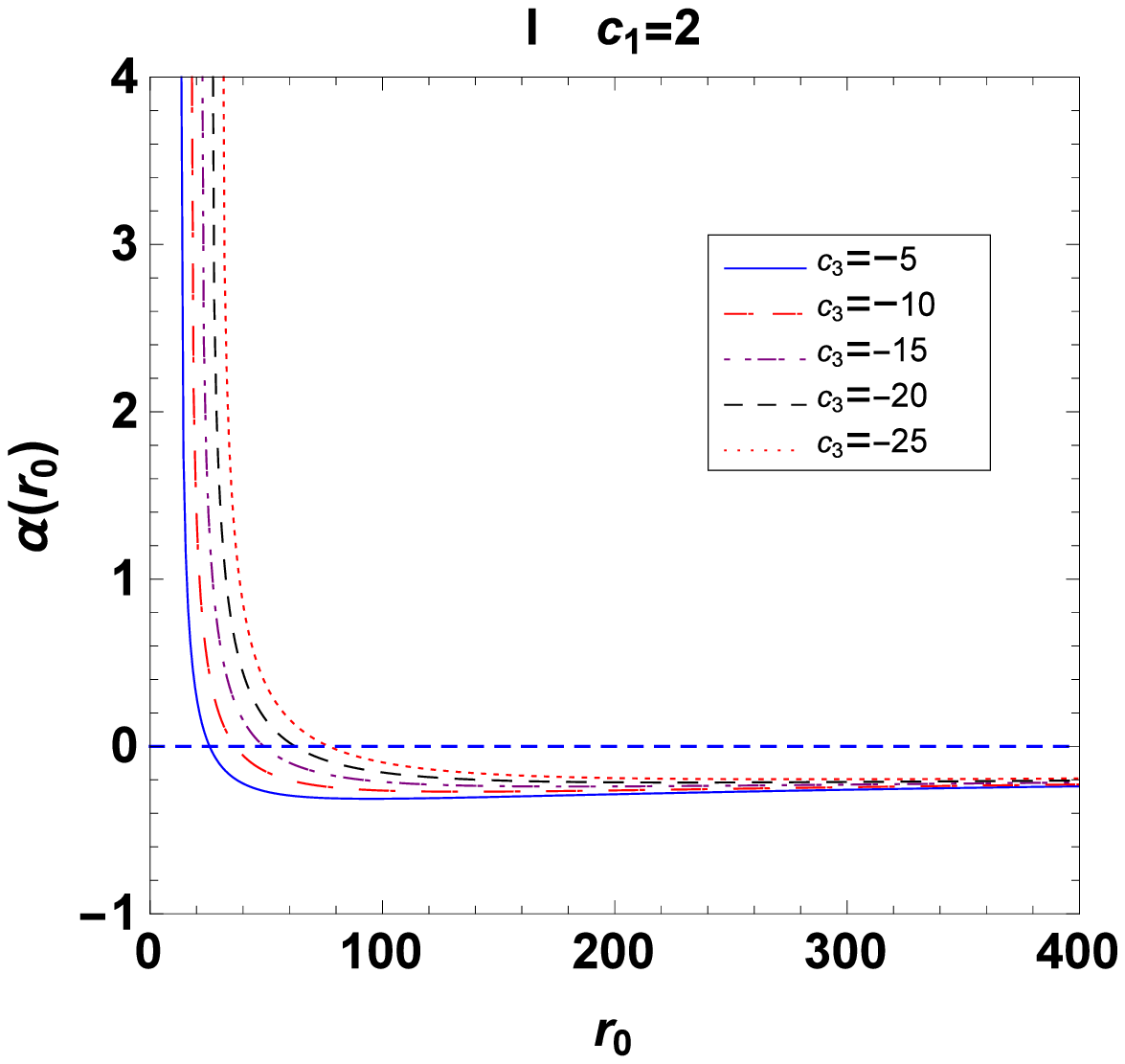}\includegraphics[width=5.2cm]{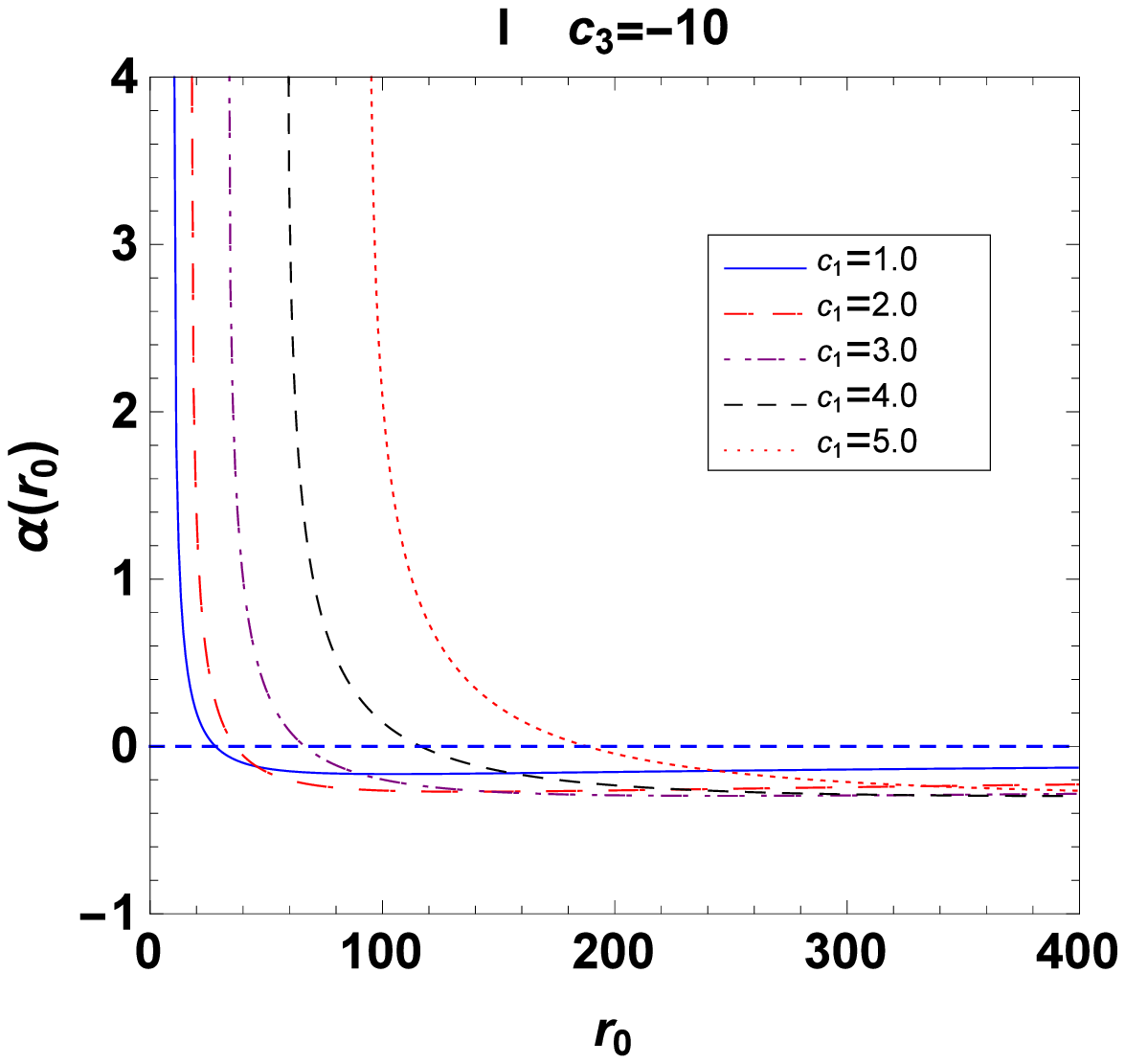}
\includegraphics[width=5.2cm]{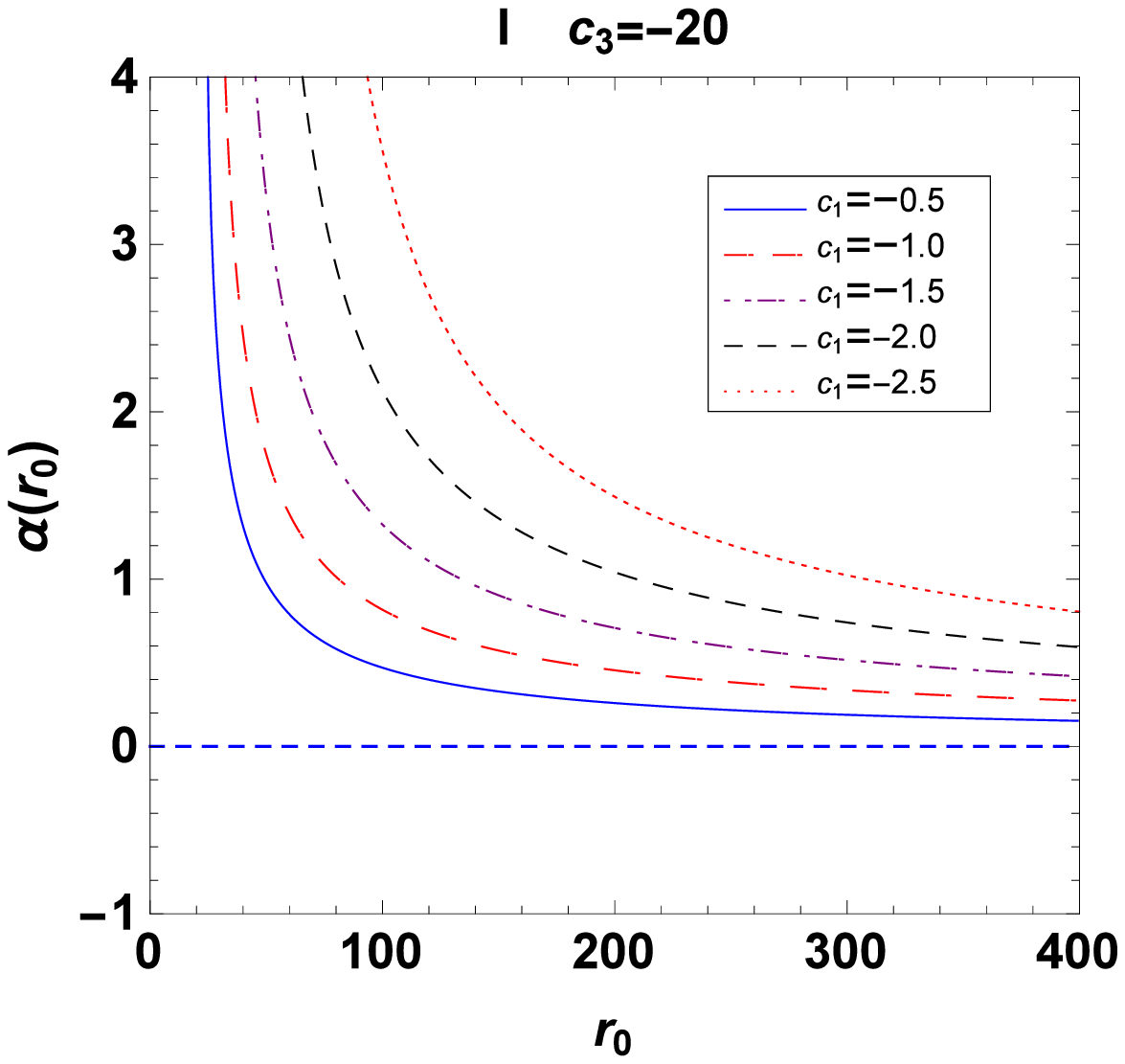}
\caption{Deflection angle $\alpha(r_0)$ as a function of the closest
distance of approach $r_0$ for the case with horizon and no photon sphere in which the parameters ($c_1, c_3$) lie in the region $I$ in Fig.(2). Here, we set $c_2=-1$.}
\end{center}
\end{figure}
In Figs.(3)-(5), we present the dependence of the deflection angle $\alpha(r_0)$ on the distance of approach $r_0$ for different parameters $c_1$ and $c_3$ in the spacetime with torsion (\ref{metr}). When there exists photon sphere for the
compact object (i.e., the parameters ($c_1, c_3$) lie in the region $II$ or $IV$ in Fig.(2) ), we find that the deflection angle possesses similar qualitative properties for the different $c_1$ and $c_2$, and it strictly increases with the decreases
of the closest distance of approach $r_{0}$ and finally becomes unlimited large as $r_{0}$ tends to the respective photon sphere radius
$r_{ps}$, i.e., $\text{lim}_{r_0\rightarrow r_{ps}}\alpha(r_{0})=\infty$ in this case. As the parameters ($c_1, c_3$)  lie in the region $III$ in Fig.(2), the singularity is naked completely since both the photon sphere and horizon vanish, we find that the deflection angle of the light
ray closing to the singularity tends to a finite value $-\pi$ for different $c_1$ and $c_2$, i.e., $\text{lim}_{r_0\rightarrow 0}\alpha(r_{0})=-\pi$, which means that the photon could not be captured by the compact object and the photon goes back along the original direction in this situation. The behavior of the deflection angle near the singularity is similar to that in the static Janis-Newman-Winicour spacetime with the naked singularity \cite{KS4,Gyulchev1}, which could be a common feature of the deflection angle near the strong naked singularity.
 As the parameters ($c_1, c_3$)  lie in the region $I$ in Fig.(2),
there exists horizon but no photon sphere, we find that the deflection angle of the light finally becomes unlimited large as $r_{0}$ tends to the respective event horizon radius
$r_{H}$, i.e., $\text{lim}_{r_0\rightarrow r_{H}}\alpha(r_{0})=\infty$, which can be understandable since the photon falls down into event horizon and then it is captured by the black hole in this case. Moreover, from Fig.(5), it is interesting to note that in this case the deflection angle is zero as $r_{0}$ is set to certain special value. This special value increases with $c_1$ and decreases with $c_3$. In the far-field limit, we find that $\text{lim}_{r_0\rightarrow\infty}\alpha(r_{0})=0$
for all values of parameters $c_1$ and $c_3$, which is a common feature in all asymptotical flat spacetimes. However, for the case with the parameters ($c_1, c_3$) lied in the regions $I$ and $III$, as $r_0\rightarrow\infty$, the deflection angle $\alpha(r_{0})$ approaches zero from the negative side as $c_1>0$, which is different from those in the spacetimes discussed in the existing literature. In the far-field limit, the deflection angle in the spacetime with torsion (\ref{metr}) can be approximated as
\begin{eqnarray}
\alpha|_{r_0\rightarrow\infty}\simeq-\frac{6\sqrt{2}\pi^{3/2}c_1}{\Gamma[1/4]^2 \sqrt{r_0}}.
\end{eqnarray}
This means that the sign of the leading term in the deflection angle is determined by the parameter $c_1$ in the far-field limit, which is consistent with the results presented in Figs.(3)-(5). Comparing with the Schwarzschild spacetime $\alpha|_{r_0\rightarrow\infty}\simeq \frac{4M}{r_0}$, we find that the behavior of the deflection angle tending to zero is more slowly in the spacetime with torsion (\ref{metr}), which is caused by the difference in the asymptotical behaviors of these two spacetimes.

\section{Strong gravitational lensing in the spacetime with torsion}

In this section, we will focus on the strong gravitational lensing in two cases where the deflection angle is divergent near the compact object: One of them is the spacetime with photon sphere in which the parameters ($c_1, c_3$) lies in the region $II$ or $IV$; The other is the spacetime without photon sphere outside horizon in which the parameters ($c_1, c_3$) lies in the region $II$. And then compare whether the deflection angles have the same divergent forms in these two cases.

Let us firstly consider the case with photon sphere outside horizon.
Following the method developed by Bozza
\cite{VB1}, we define a variable
\begin{eqnarray}
z=1-\frac{r_0}{r},
\end{eqnarray}
and then the integral (\ref{int1}) becomes
\begin{eqnarray}
I(r_0)=\int^{1}_{0}R_1(z,r_0)f_1(z,r_0)dz,\label{in1}
\end{eqnarray}
where
\begin{eqnarray}
R_1(z,r_0)&=&2\sqrt{A(r)B(r)C(r)},
\end{eqnarray}
\begin{eqnarray}
f_1(z,r_0)&=&\frac{1}{\sqrt{A(r_0)C(r)-A(r)C(r_0)}}.
\end{eqnarray}
After some analysis, one can find that the function $R_1(z, r_0)$ is regular for all values of $z$ and $r_0$, but $f_1(z, r_0)$ diverges as $z$ tends to zero. This means that
the integral (\ref{in1}) can be split into the divergent part $I_{D_1}(r_0)$ and the regular
part $I_{R_1}(r_0)$, i.e.,
\begin{eqnarray}
I_{D_1}(r_0)&=&\int^{1}_{0}R_{1}(0,r_{ps})f_{10}(z,r_0)dz, \nonumber\\
I_{R_1}(r_0)&=&\int^{1}_{0}[R_1(z,r_0)f_1(z,r_0)-R_1(0,r_{ps})f_{10}(z,r_0)]dz
\label{intbr},
\end{eqnarray}
Expanding the argument of the square
root in $f_1(z,r_0)$ to the second order in $z$, one can obtain the
function $f_{10}(z,r_0)$,
\begin{eqnarray}
f_{10}(z,r_0)=\frac{1}{\sqrt{p_1(r_0)z+q_1(r_0)z^2}},
\end{eqnarray}
with
\begin{eqnarray}
p_1(r_0)&=& 2r_0(\sqrt{r_0}+2c_1)(\sqrt{r_0}+3c_1),  \nonumber\\
q_1(r_0)&=&\frac{r_0}{2}(3\sqrt{r_0}+8c_1)(2\sqrt{r_0}+3c_1).
\end{eqnarray}
For the case with photon sphere outside horizon, we have $c_1<0$. Thus, when $r_0$ is equal to the radius of photon sphere $r_{ps}$, i.e., $\sqrt{r_{ps}}=-3c_1$, the
coefficient $p_1(r_0)$ vanishes and the leading term of the
divergence in $f_0(z,r_0)$ is $z^{-1}$, and then the deflection angle can be
expanded in the form \cite{VB1}
\begin{eqnarray}
\alpha(\theta)=-\bar{a}_1\ln{\bigg(\frac{\theta
D_{OL}}{u_{ps}}-1\bigg)}+\bar{b}_1+O(u-u_{ps}),
\end{eqnarray}
with
\begin{eqnarray}
&\bar{a}_1&=\frac{R_1(0,r_{ps})}{2\sqrt{q_1(r_{ps})}}=
-2\sqrt{3}c_1\bigg[2(6c^2_1+c_3)-3c^2_1\ln(-3c_1)\bigg]^{-1/2}, \nonumber\\
&\bar{b}_1&=
-\pi+b_{R_1}+\bar{a}_1\ln{\frac{r^2_{ps}[C''(r_{ps})A(r_{ps})-C(r_{ps})A''(r_{ps})]}{u_{ps}
\sqrt{A^3(r_{ps})C(r_{ps})}}}, \nonumber\\
&b_{R_1}&=I_{R_1}(r_{ps}), \;\;\;\;\;u_{ps}=\frac{r_{ps}}{\sqrt{A(r_{ps})}},
\end{eqnarray}
which means that the deflection angle
diverges logarithmically as the light is close to the photon sphere. Here
$D_{OL}$ denotes the distance between observer and gravitational
lens, $\bar{a}$ and $\bar{b}$ are the strong field limit
coefficients which depend on the metric functions evaluated at
$r_{ps}$.
\begin{figure}[ht]\label{pas10}
\begin{center}
\includegraphics[width=6cm]{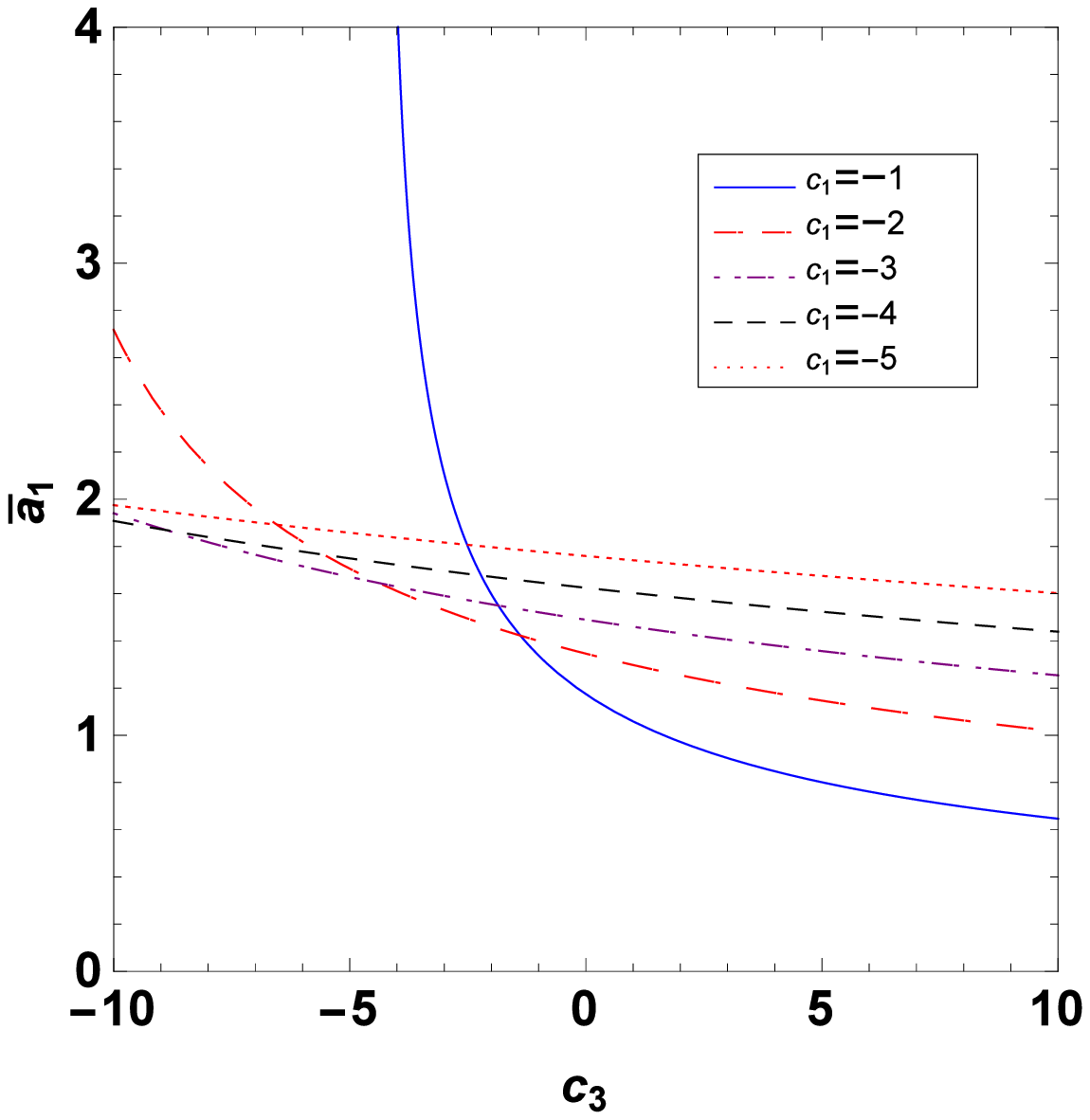}\includegraphics[width=6cm]{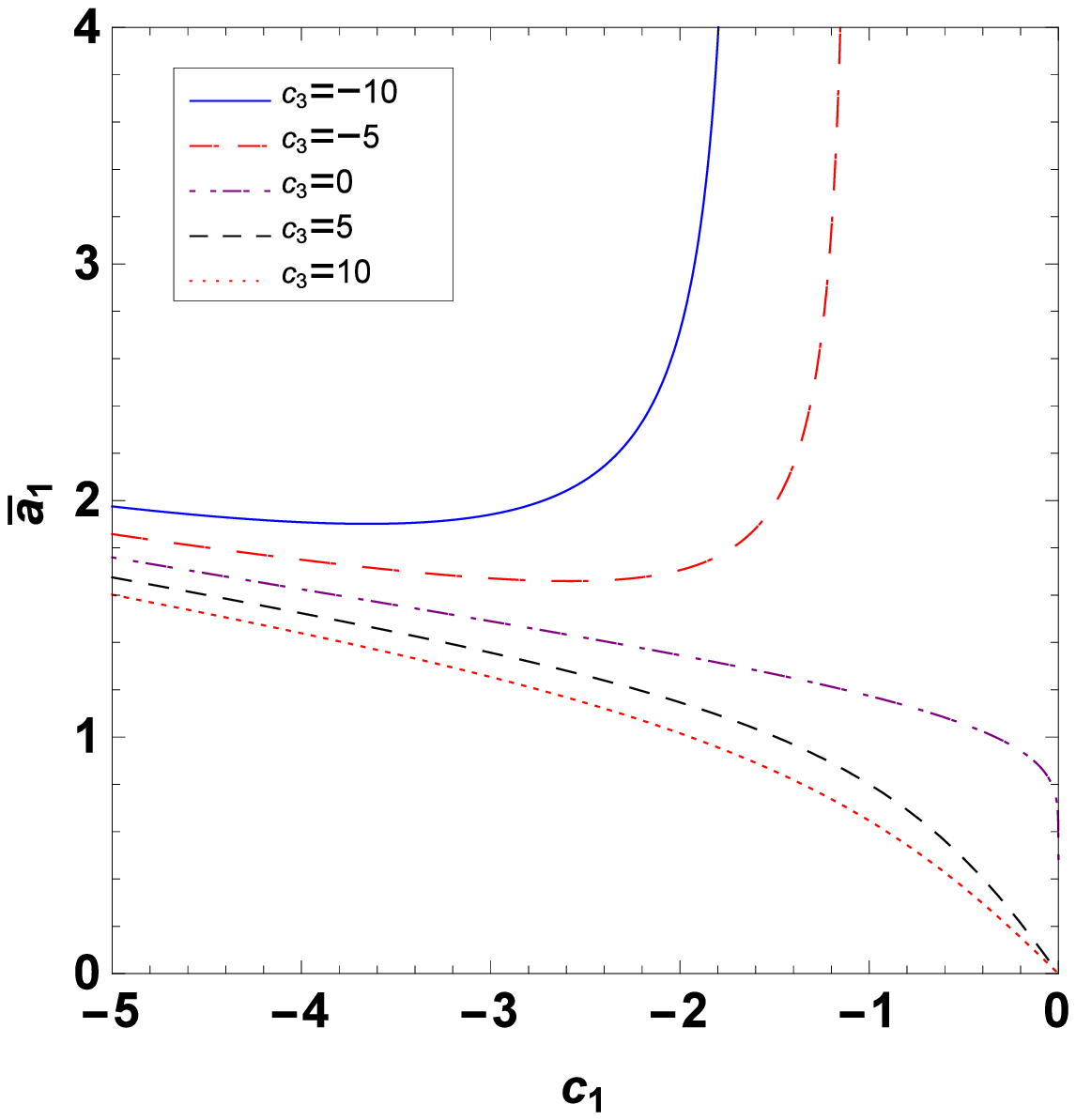}\\
\includegraphics[width=6cm]{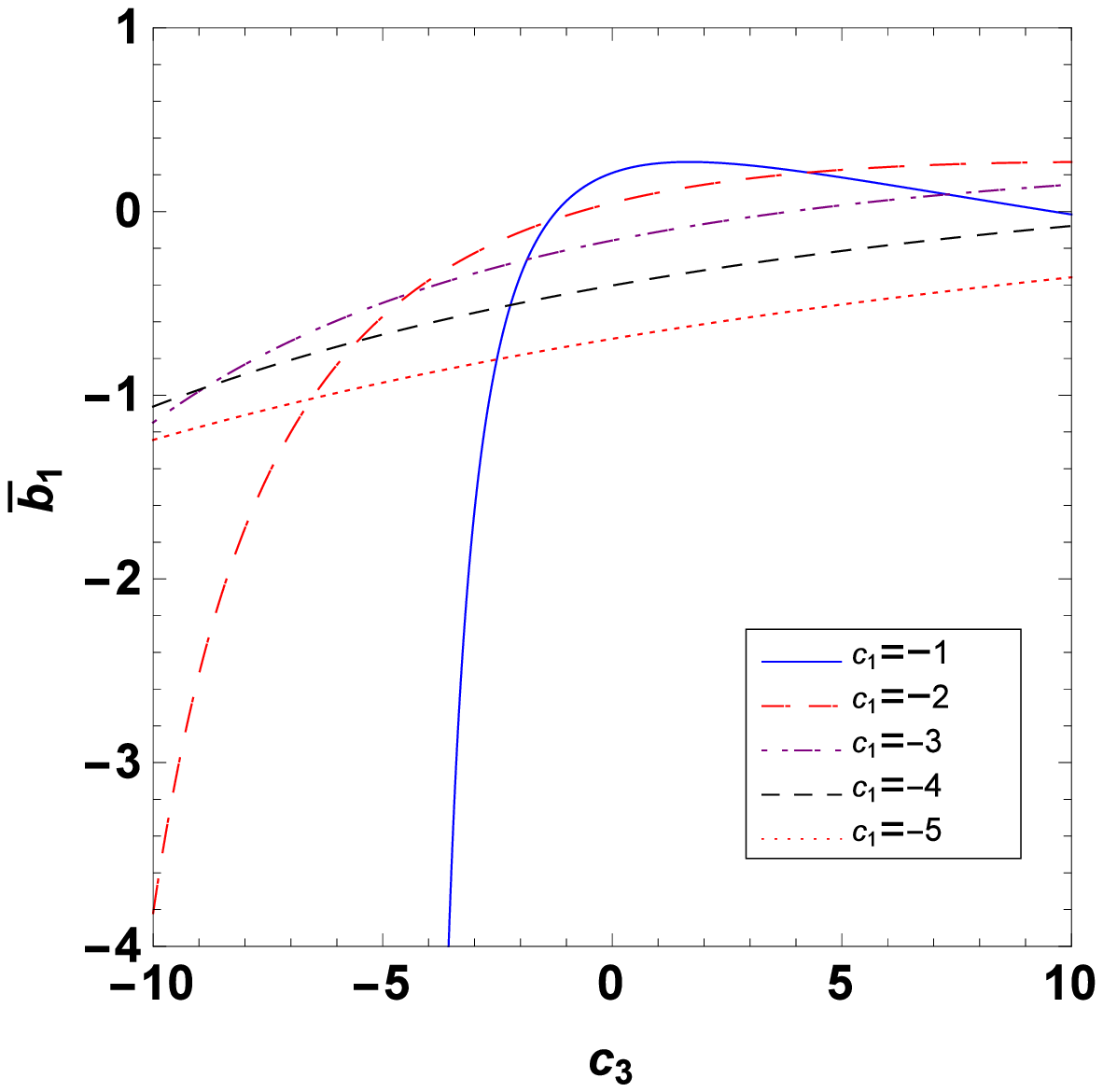}\includegraphics[width=6cm]{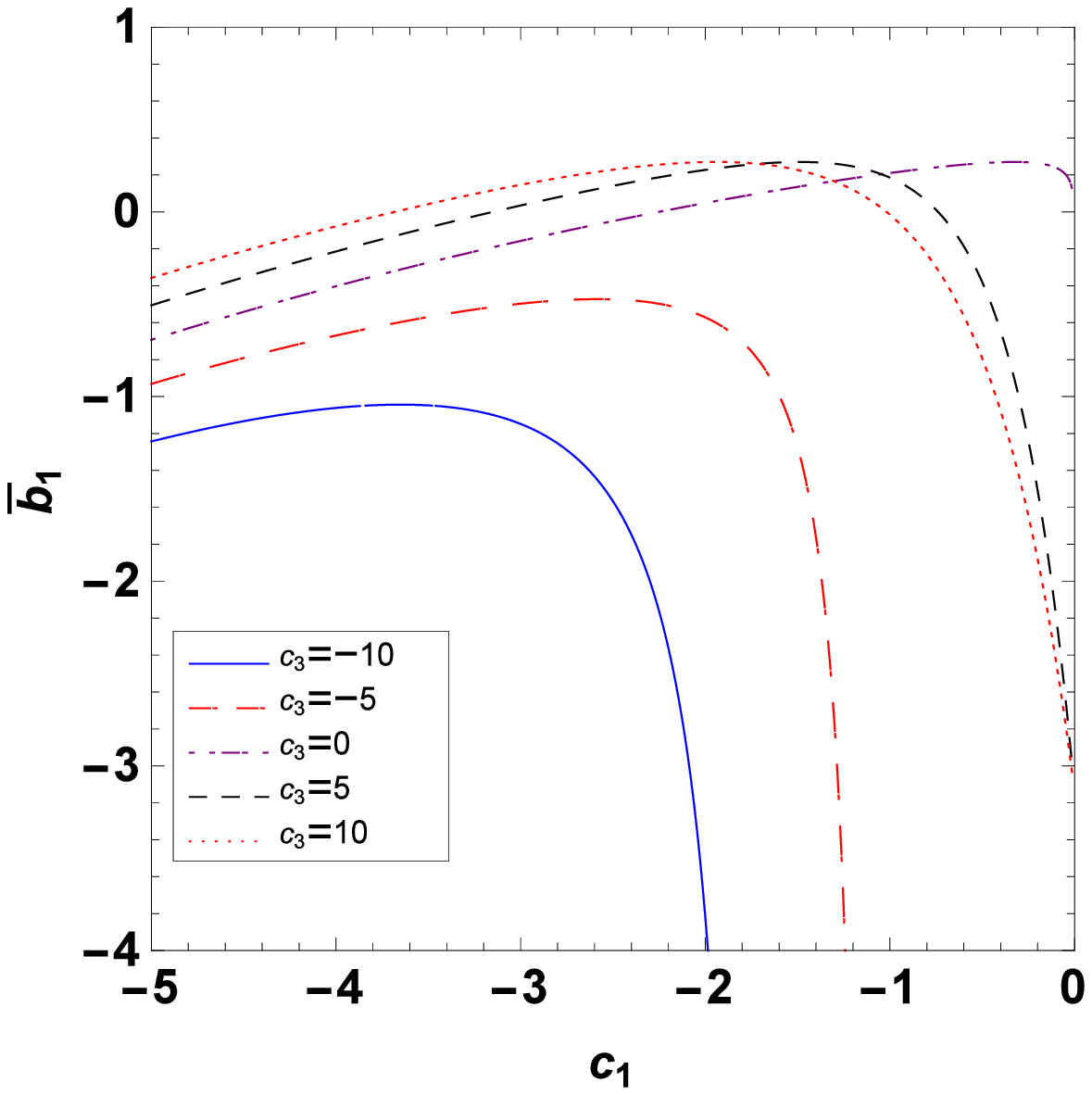}
\caption{Change of the strong deflection limit coefficients $\bar{a}_1$ and $\bar{b}_1$ with the parameters $c_1$ and $c_3$ in the spacetime with torsion (\ref{metr}). Here, we set $c_2=-1$.}
\end{center}
\end{figure}
The changes of the coefficients ($\bar{a}_1$ and $\bar{b}_1$ ) with the parameters $c_1$ and $c_3$ is shown in Fig.(6). For the fixed $c_1$, the coefficient $\bar{a}_1$ decreases monotonously with $c_3$. The coefficient $\bar{b}_1$ increases with $c_3$ for the larger absolute value of $c_1$, but first increases and then decreases for the smaller one. For the fixed $c_3$, the coefficient $\bar{a}_1$ increases with $c_1$ for the negative $c_3$ and decreases for the positive one. The coefficient $\bar{b}_1$ first increases and then decreases
with $c_1$ for all $c_3$ in this case.

We are now in position to study the case without photon sphere outside horizon. Considering the divergence of the deflection angle for the light near the horizon in this case, we rewrite the functions $R_2(z, r_0)$ and $f_2(z, r_0)$ in (\ref{in1}) as
\begin{eqnarray}
R_2(z,r_0)&=&2\sqrt{A(r)C(r)},
\end{eqnarray}
\begin{eqnarray}
f_2(z,r_0)&=&\frac{\sqrt{B(r)}}{\sqrt{A(r_0)C(r)-A(r)C(r_0)}}.
\end{eqnarray}
which ensures that the function $R_2(z, r_0)$ is regular for all values of $z$ and $r_0$ and only the function $f_2(z, r_0)$ diverges as $z$ tends to zero. Correspondingly,
the divergent part $I_{D_2}(r_0)$ and the regular
part $I_{R_2}(r_0)$ in (\ref{in1}) become
\begin{eqnarray}
I_{D_2}(r_0)&=&\int^{1}_{0}R_2(0,r_{H})f_{20}(z,r_0)dz, \nonumber\\
I_{R_2}(r_0)&=&\int^{1}_{0}[R_2(z,r_0)f_2(z,r_0)-R_2(0,r_{H})f_{20}(z,r_0)]dz
\label{intbr2}.
\end{eqnarray}
Similarly, expanding the argument of the square
root in $f_2(z,r_0)$ to the second order in $z$, one can obtain the
function $f_{20}(z,r_0)$,
\begin{eqnarray}
f_{20}(z,r_0)=\frac{1}{\sqrt{p_2(r_0)z+q_2(r_0)z^2}},
\end{eqnarray}
with
\begin{eqnarray}
p_2(r_0)&=& (\sqrt{r_0}+2c_1)(\sqrt{r_0}+3c_1)\bigg[
2(r_0+c_1\sqrt{r_0}+c_3)-3c^2_1\ln(2\sqrt{r_0}+3c_1)\bigg],  \nonumber\\
q_2(r_0)&=&\sqrt{r_0}\bigg\{\frac{4(\sqrt{r_0}+2c_1)^2(\sqrt{r_0}+3c_1)}{2\sqrt{r_0}+3c_1}
+\frac{(2\sqrt{r_0}+5c_1)}{4}\bigg[
2(r_0+c_1\sqrt{r_0}+c_3)-3c^2_1\ln(2\sqrt{r_0}+3c_1)\bigg]\bigg\}.
\end{eqnarray}
When $r_0$ is equal to the radius of event horizon $r_{H}$, the
coefficient $p_2(r_0)$ vanishes and the leading term of the
divergence in $f_{20}(z,r_0)$ is $z^{-1}$. This means that the deflection angle diverges also logarithmically as the light is close to the event horizon. The form the deflection angle near the event horizon can further be approximated as
\begin{eqnarray}
\alpha(\theta)=-\bar{a}_2\ln{\bigg(\frac{\theta
D_{OL}}{u_{H}}-1\bigg)}+\bar{b}_2+O(u-u_{H}),
\end{eqnarray}
with
\begin{eqnarray}
&\bar{a}_2&=\frac{R_2(0,H)}{\sqrt{q_2(r_{H})}}=
\bigg[\frac{2\sqrt{r_H}+3c_1}{\sqrt{r_H}+3c_1}\bigg]^{1/2}, \nonumber\\
&\bar{b}_2&=
-\pi+b_{R_2}+\bar{a}_2\ln{\frac{2r_H[C'(r_{H})A(r_{H})-C(r_{H})A'(r_{H})]}{
u_{H}\sqrt{A^3(r_{H})C(r_{H})}}}, \nonumber\\
&b_{R_2}&=I_{R_2}(r_{H}), \;\;\;\;\;u_{H}=\frac{r_{H}}{\sqrt{A(r_{H})}}.
\end{eqnarray}
Comparing with the previous cases with photon sphere, one can find that although the deflection angle in the strong field limit own the same forms, the coefficients in the strong gravitational lensing formulae are different in these two cases.
\begin{figure}[ht]\label{pas20}
\begin{center}
\includegraphics[width=5.2cm]{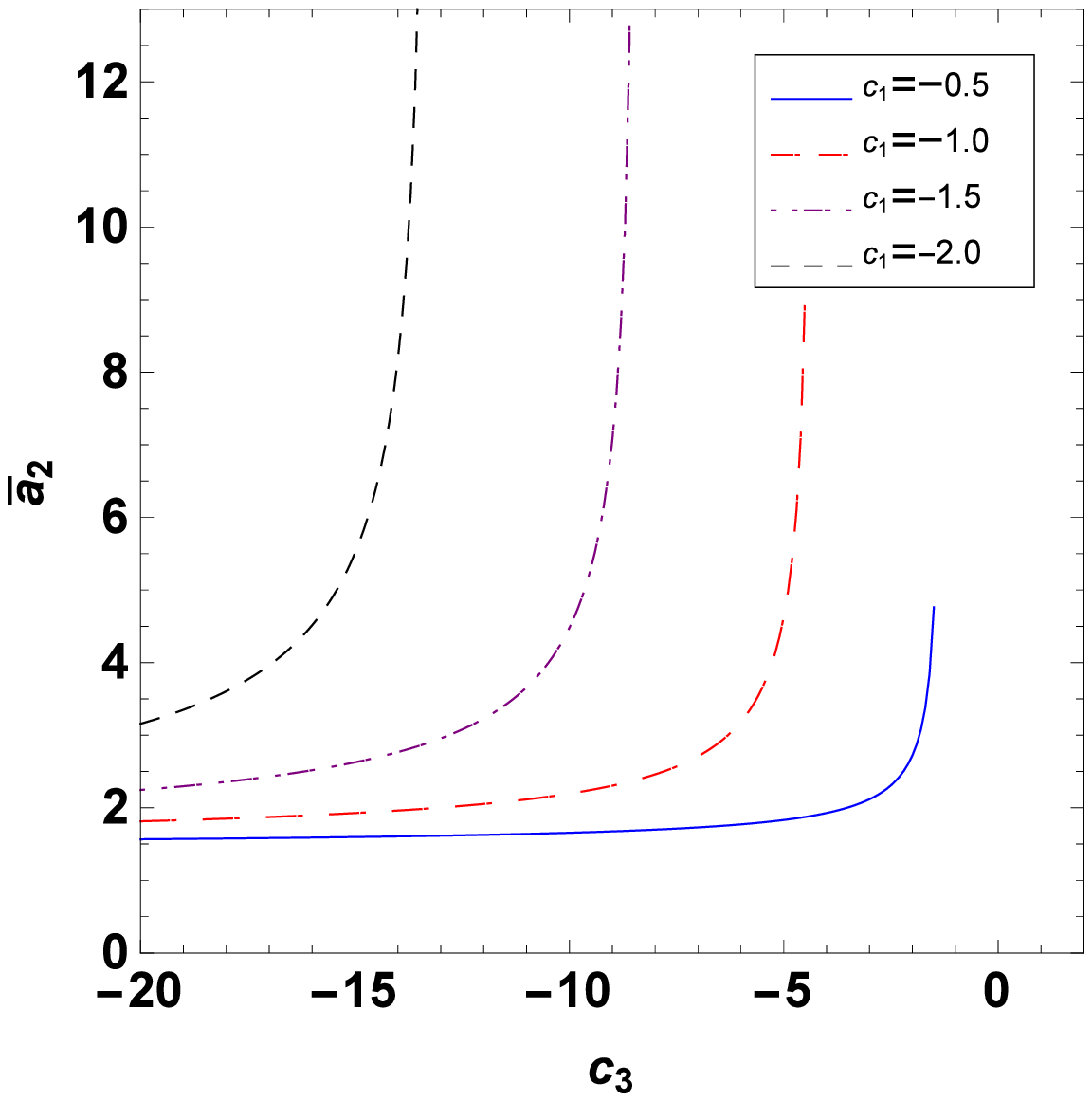}\includegraphics[width=5.3cm]{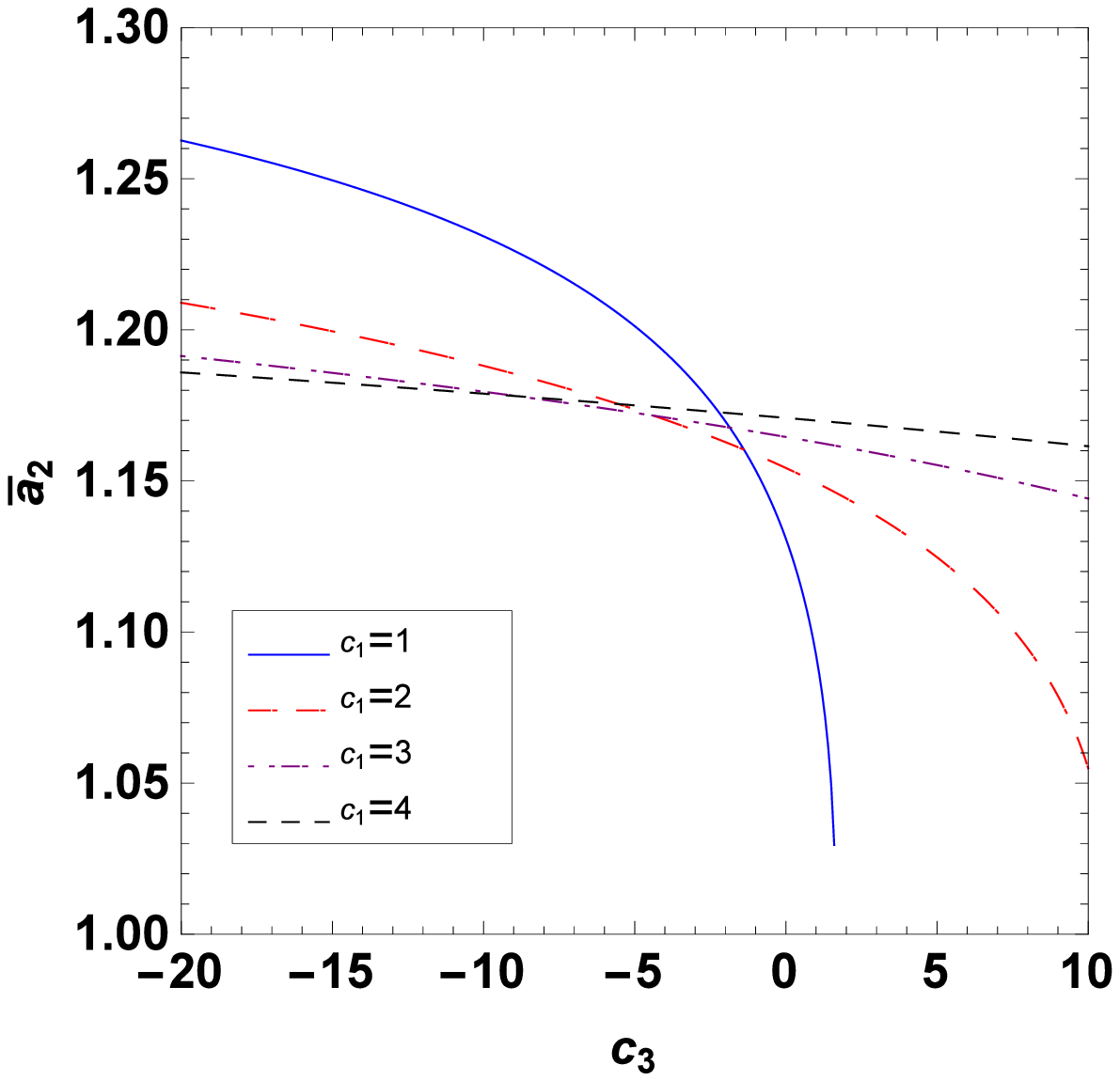}
\includegraphics[width=5.2cm]{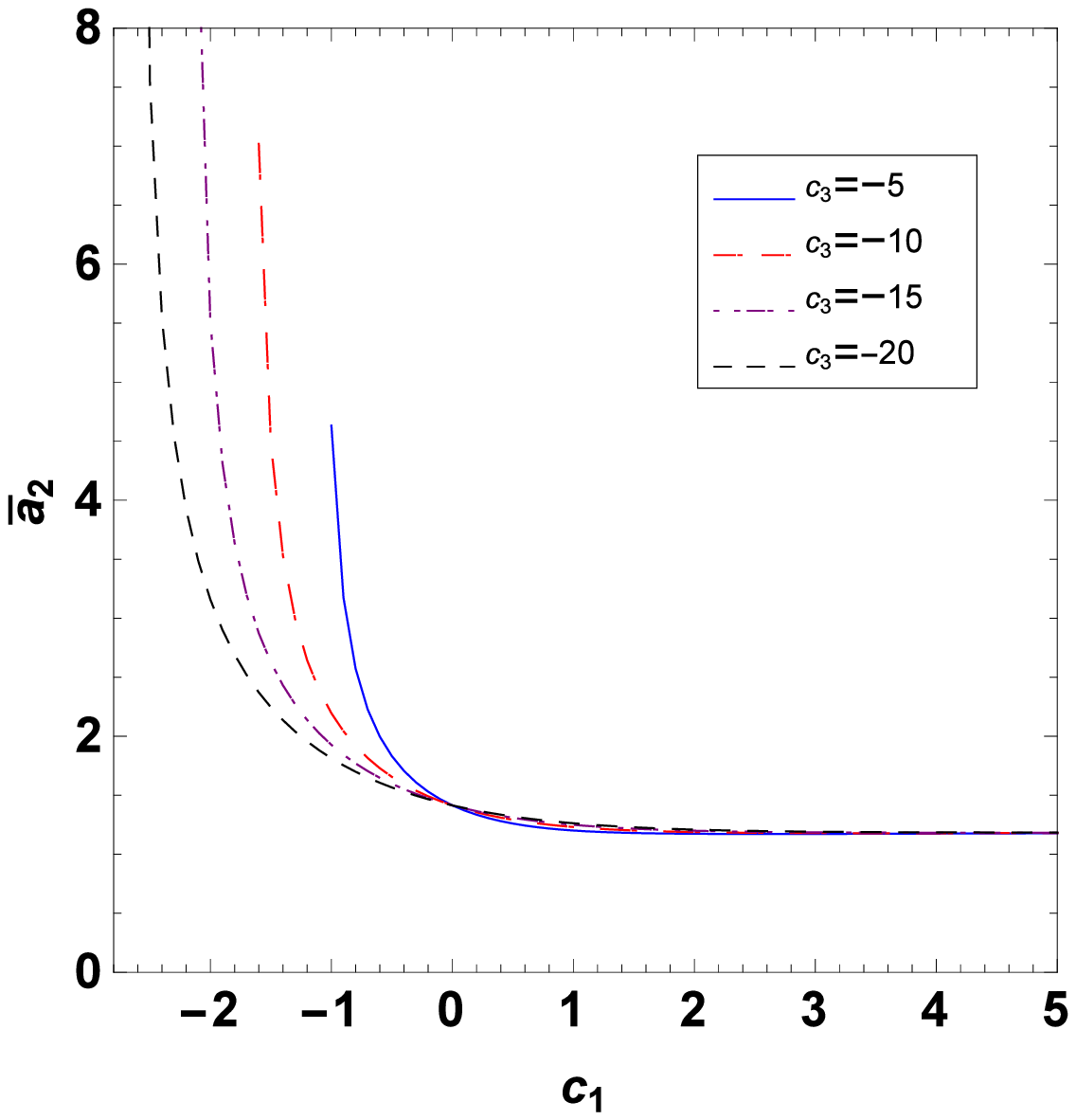}\\
\includegraphics[width=5.2cm]{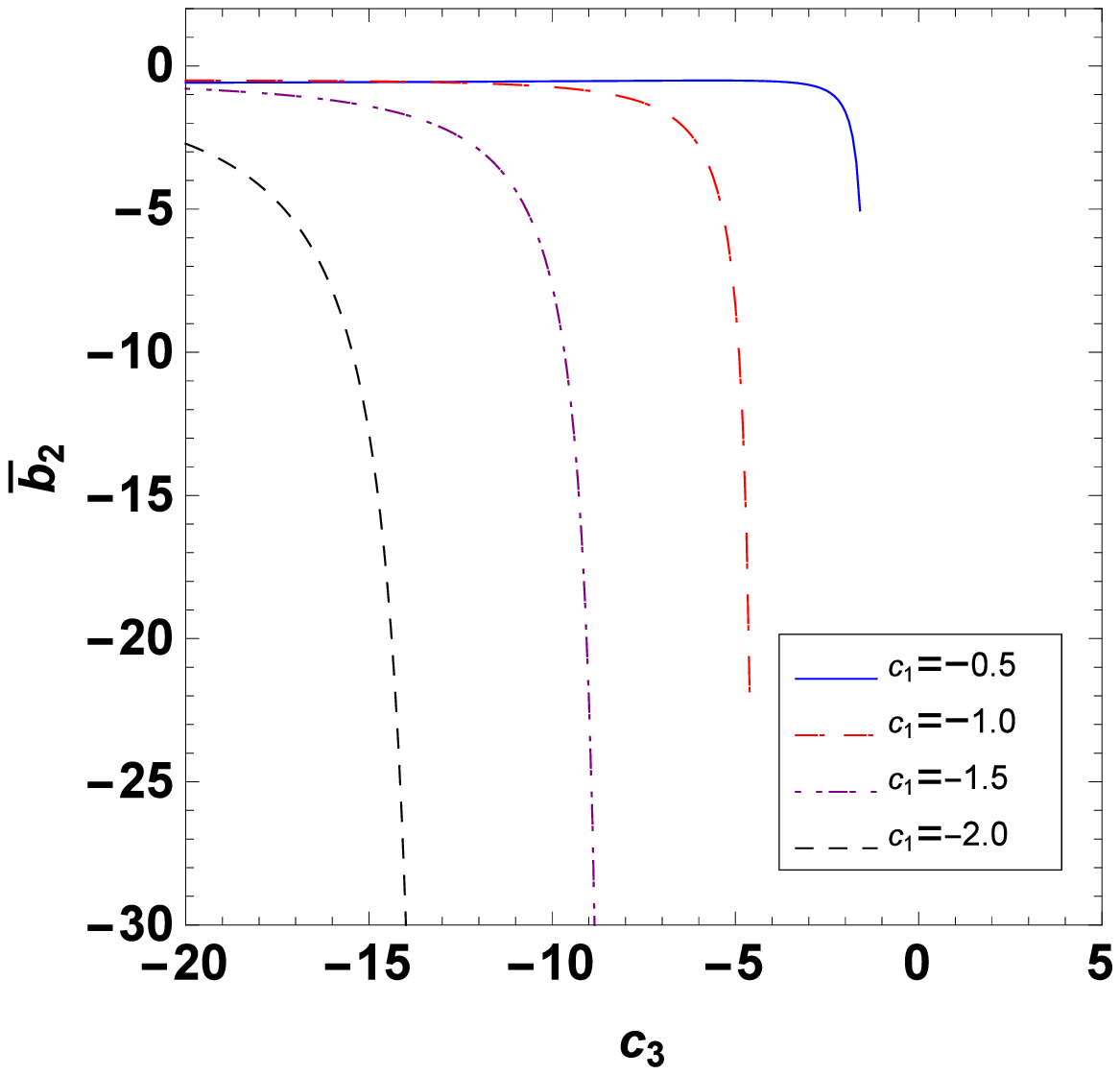}\includegraphics[width=5.3cm]{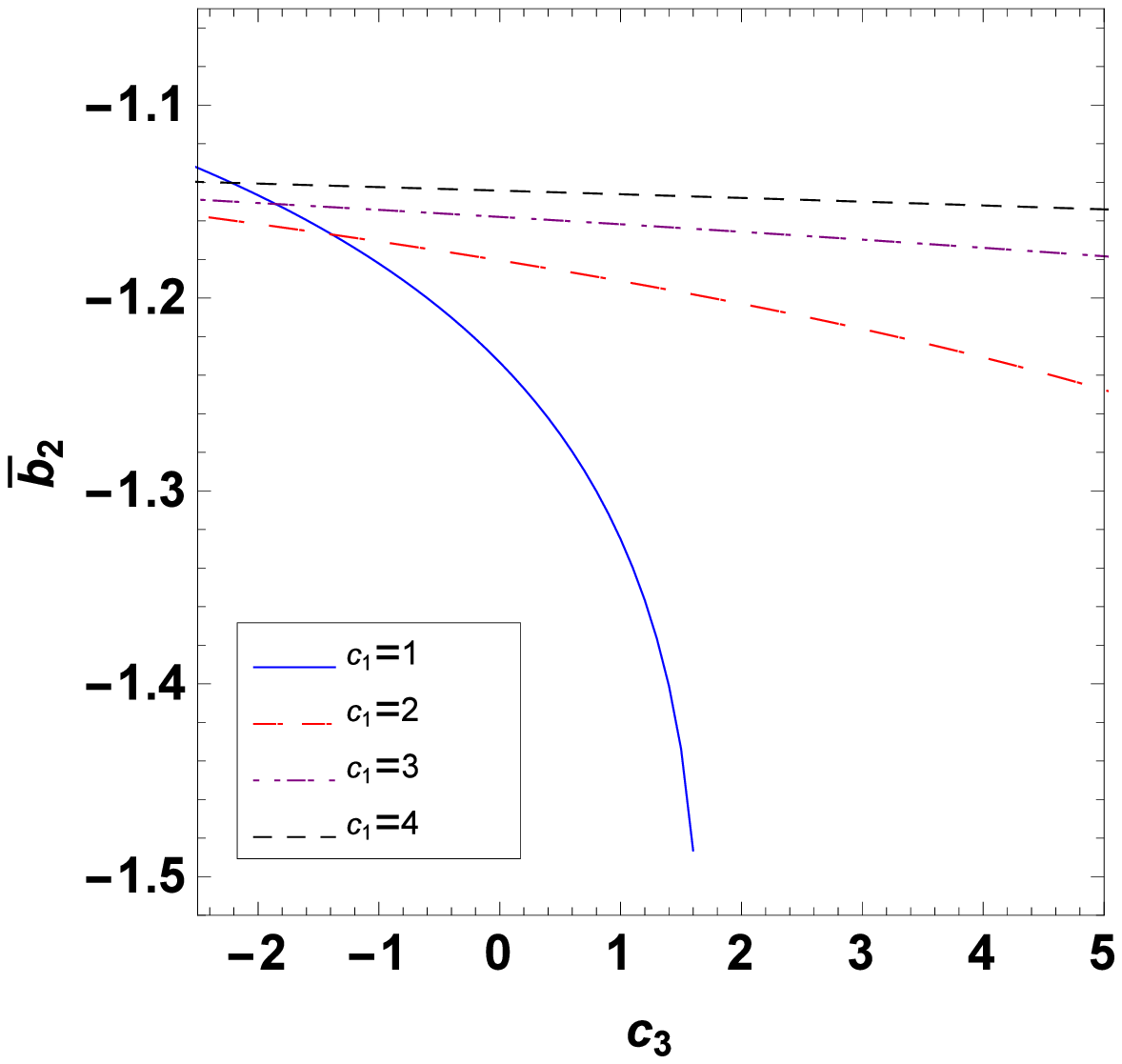}
\includegraphics[width=5.2cm]{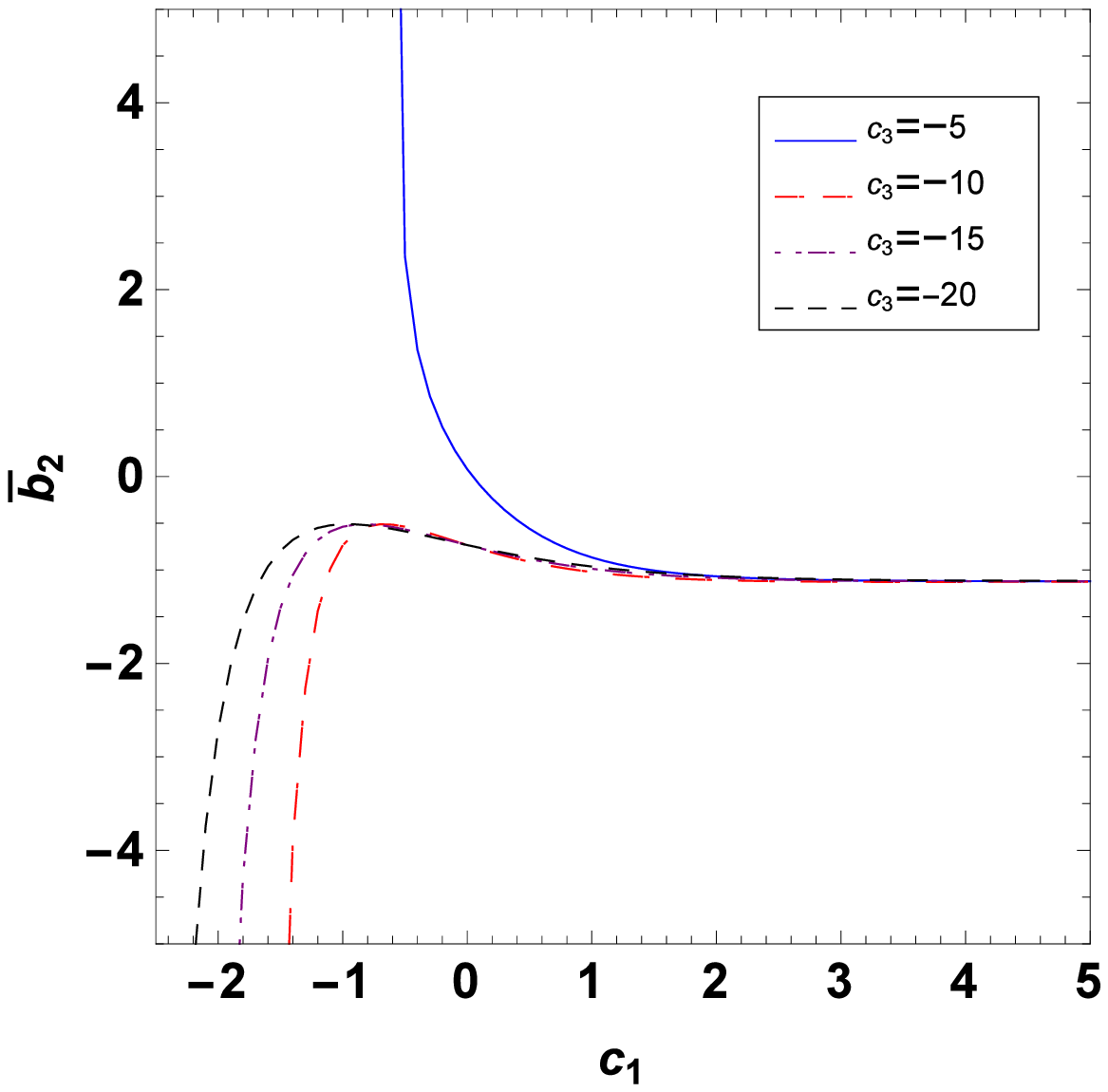}
\caption{Change of the strong deflection limit coefficients $\bar{a}_2$ and $\bar{b}_2$ with the parameters $c_1$ and $c_3$ in the spacetime with torsion (\ref{metr}). Here, we set $c_2=-1$.}
\end{center}
\end{figure}
We also plot the changes of the coefficients ($\bar{a}_2$ and $\bar{b}_2$ ) with the parameters $c_1$ and $c_3$ is shown in Fig.(7). For the fixed $c_1$, the coefficient $\bar{a}_2$ increases with $c_3$ for the negative $c_1$ and decreases for the positive $c_1$.
The coefficient $\bar{b}_2$ increases with $c_3$ for all $c_1$.  For the fixed $c_3$,
$\bar{a}_2$ decreases with $c_1$ and tends to a constant $\frac{2\sqrt{3}}{3}$ for the case with a larger positive $c_1$. The change of $\bar{b}_2$ with $c_1$ becomes more complicated.
For the case with positive $c_3$ or with the smaller absolute value of negative $c_3$,
$\bar{b}_2$ decreases with $c_1$.  For the case with the larger absolute value of negative $c_3$, it first increases and then decreases with $c_1$. Similarly, in the case with a larger positive $c_1$, $\bar{b}_2$ reduces only a function of $c_1$ and is independent of the parameter $c_3$.

\section{summary}

In this paper we have investigated the propagation of photon in a spherically symmetric spacetime with torsion in the generalized ECKS theory of gravity. We find that the torsion parameters change the spacetime structure which imprint in the photon sphere, the deflection angle and the strong gravitational
lensing. The condition of existence of horizons is not inconsistent with that of the photon sphere and  the boundaries of the existence of horizon and of the photon sphere split the whole region into four regions in the panel ($c_1, c_3$). In the cases with photon sphere,
the deflection angle of the light ray near the photon sphere diverges logarithmically in the strong-field limit as in the usual spacetime of a black hole or a weak naked singularity.
In the strong naked singularity case where there is no photon sphere and no horizon,
the deflection angle of the light ray closing very to the singularity tends to a fixed value $-\pi$, which is independent of torsion parameters $c_1$ and $c_3$.  This behavior is similar to that in the static Janis-Newman-Winicour spacetime \cite{KS4,Gyulchev1}, which could be a common feature of the deflection angle near the static strong naked singularity.
Especially,
there exists a novel case in which there is horizon but no photon sphere for the spacetime (\ref{metr}). In this special case, we find that the deflection angle of the light ray near the event horizon also diverges logarithmically, but the coefficients in the strong-field limit are different from those in the case with photon sphere. In the far-field limit,  the deflection angle $\text{lim}_{r_0\rightarrow\infty}\alpha(r_{0})=0$
for all values of parameters $c_1$ and $c_3$, which is a common feature in all asymptotical flat spacetimes. However, for the case with the positive $c_1$, we find that the deflection angle $\alpha(r_{0})$ approaches zero from the negative side as $r_0\rightarrow\infty$, which is different from those in the usual spacetimes.

\section{\bf Acknowledgments}

This work was partially supported by the Scientific Research
Fund of Hunan Provincial Education Department Grant
No. 17A124. J. Jing's work was partially supported by
the National Natural Science Foundation of China under
Grant No. 11475061.

\end{document}